%% file: 00_main.tex
\documentclass[conference, 10pt]{IEEEtran}
\ifCLASSINFOpdf
\else
\fi
\usepackage{amsmath}

\usepackage{algorithm}
\usepackage{tabularx}
\usepackage{algorithmic}
\usepackage{enumitem}
\usepackage{xcolor}
\usepackage{graphicx}
\usepackage{multirow}
\usepackage[table]{colortbl}
\usepackage{rotating}
\usepackage{caption}
\usepackage{subcaption}
\PassOptionsToPackage{hyphens}{url}\usepackage{hyperref}

\begin{document}
%
\title{Towards Fairness-aware Crowd Management System and Surge Prevention in Smart Cities}

\author{\IEEEauthorblockN{Yixin Zhang\textsuperscript{\textsection}}
\IEEEauthorblockA{Carnegie Mellon University\\
yixinzha@cs.cmu.edu}
\and
\IEEEauthorblockN{Tianyu Zhao}
\IEEEauthorblockA{University of California, Irvine\\
tzhao15@uci.edu}
\and
\IEEEauthorblockN{Salma Elmalaki}
\IEEEauthorblockA{University of California, Irvine\\
salma.elmalaki@uci.edu}}


%


\maketitle

\begingroup\renewcommand\thefootnote{\textsection}
\footnotetext{This work was done when the author was at UC, Irvine.}
\endgroup

\thispagestyle{plain}
\pagestyle{plain}

\begin{abstract}
Instances of casualties resulting from large crowds persist, highlighting the existing limitations of current crowd management practices in Smart Cities. One notable drawback is the insufficient provision for disadvantaged individuals who may require additional time to evacuate due to their slower running speed. Moreover, the existing escape strategies may fall short of ensuring the safety of all individuals during a crowd surge. To address these pressing concerns, this paper proposes two crowd management methodologies. Firstly, we advocate for implementing a fair evacuation strategy following a surge event, which considers the diverse needs of all individuals, ensuring inclusivity and mitigating potential risks. Secondly, we propose a preventative approach involving the adjustment of attraction locations and switching between stage performances in large-crowded events to minimize the occurrence of surges and enhance crowd dispersion. We used high-fidelity crowd management simulators to assess the effectiveness of our proposals. Our findings demonstrate the positive impact of the fair evacuation strategy on safety measures and inclusivity, which increases fairness by $41.8\%$ on average. Furthermore, adjusting attraction locations and stage performances has shown a significant reduction in surges by $34\%$ on average, enhancing overall crowd safety.

\end{abstract}


%
\IEEEpeerreviewmaketitle

\input{01_introduction}

\input{02_background}

\input{03_framework}
\input{04_fairness_vadere}

\input{05_preventive_netlogo}

\input{06_conclusion}

\section*{Acknowledgment}

This research was partially supported by NSF awards \# 2105084 and \# 2339266.

\IEEEtriggercmd{\enlargethispage{-5in}}


\bibliographystyle{IEEEtranS}
\bibliography{IEEEabrv,mybibfile}
%



\input{07_appendix}

\end{document}

%% file: 01_introduction.tex
\vspace{-2mm}
\section{Introduction}

In the contemporary urban landscape, managing crowd dynamics in confined spaces has emerged as a pivotal concern for ensuring public safety within smart cities~\cite{illiyas2013human}. The tragic incidents at Itaewon and the Astroworld Festival underscore the urgency of advancing crowd management techniques to prevent similar catastrophes. The Itaewon disaster~\cite{sharma2023global}, where a dense crowd led to 156 fatalities and 170 injuries, and the Astroworld Festival~\cite{astroworld}, which saw 10 deaths and numerous injuries due to a surge near the stage, highlight the critical challenges in crowd control during mass gatherings. These events bring to light the complexities of managing large groups, particularly when faced with limited entry points, uneven terrains, and unforeseen bottlenecks that exacerbate the risk of crowd-induced accidents. In response to these challenges, the concept of ``Smart Crowd Management and Control Systems'' (CMS) within the framework Cyber-Physical Systems (CPS) for smart cities presents a holistic approach to overseeing large crowds. CMS is tasked with monitoring, directing, and managing large groups of people---with an eye toward safety, efficiency, and satisfaction.
CMS requires a diverse range of knowledge, including engineering, technology, and understanding of crowd behavior~\cite{sharma2018review}. 
Effective crowd management involves multiple stages: pre-event planning, event monitoring and control, post-event feedback, and improvement. This holistic approach ensures continuous enhancement in crowd management strategies for future events. 
However, we believe that each of these approaches has different effects on healthy individuals, people with disabilities, children, pregnant women, and neurodivergent communities. Therefore, while managing the crowd flow, CMS should not only aim to maximize the efficiency, but also ensure fairness in distributing the adverse effect of the CMS control actions among different individuals. %
%
%
%
This vision is attainable with the advancement in sensor technologies to estimate the human state through wearable devices, and the decision-making algorithms that provide trade-offs between system performance, fairness~\cite{zhao2024fairo, zhao2024fina, elmalaki2021fair}, and privacy in multi-human environment~\cite{taherisadr2023adaparl, taherisadr2024hilt}. This technological leap in human sensing and decision-making algorithms should be exploited in CMS to ensure inclusivity in crowd management algorithms.




\textbf{In this paper, we advocate for implementing a fair evacuation strategy and prevention approaches that account for the diverse needs of all individuals. By embracing an inclusive approach, we can provide the necessary time and assistance to disadvantaged individuals, helping them to evacuate safely and efficiently. Through thoughtful planning and coordination, we can mitigate potential risks and minimize casualties.} The contribution of this paper can be summarized as follows:
\begin{itemize}[topsep=0pt, leftmargin=*]
\item \textbf{Fair evacuation strategies:} Proposing evacuation routes designed to serve diverse groups within various crowd-gathering contexts, aiming for efficiency and fairness, especially for vulnerable populations. We introduce a metric, ``Normalized Evacuation Time Disparity,'' to evaluate and compare different crowd scenarios and evacuation strategies.
\item \textbf{Surge prevention:}  Suggesting a preventative approach that involves alternating stage performances at specific intervals to control crowd density and distribution effectively. This method utilizes metrics like Panic State, Surge State, and Crowd State to dynamically assess and manage crowd conditions, thereby optimizing event scheduling to enhance crowd management overall.
\item \textbf{Simulation and Validation:} We utilize high-fidelity Crowd Management simulators, including Vadere~\cite{kleinmeier2019vadere} and NetLogo~\cite{tisue2004netlogo} to simulate various crowd scenarios and refining our proposed strategies.
\end{itemize}

%% file: 02_background.tex
\vspace{-2mm}
\section{Background \& Related Work}\vspace{-2mm}
Crowd surges pose a significant risk when a large number of individuals attempt to enter or exit a confined area, leading to increased pressure and potential danger. In tightly packed crowds, people lose control of their movements and face difficulties breathing. The lack of space for recovery makes stumbling or falling particularly hazardous, putting individuals at risk of suffocation and injuries from being crushed. The probability of a surge occurring is closely related to crowd density. When there are 4-5 people per square meter, the crowd remains relatively safe, with enough space for individuals to make movement decisions. However, when the number exceeds 6 per square meter, the limited available space forces tight packing and diminishes individual control, significantly increasing the likelihood of a surge~\cite{abuarafah2012real}. A single stumble or jolt within the crowd can trigger a chain reaction, creating voids that disrupt the crowd's equilibrium. Subsequently, more people stumble into these voids, setting off a domino effect and generating additional voids. This interplay of forces can cause abrupt collapses, intensifying pressure and chaos within the crowd, potentially leading to injuries or fatalities if not managed properly~\cite{aalami2020fairness}. Throughout history, large-scale stampedes have taken place worldwide, leading to severe loss of life and property damage. On October 29, 2022, a Halloween event occurred in Seoul, South Korea, attracting tens of thousands of costumed attendees to the Itaewon district. This marked the first unrestricted Halloween celebration in over two years due to COVID-19 lockdowns. The massive crowd in the narrow streets, coupled with limited entry and exit points, created a dangerous situation. Videos from that night show trapped individuals struggling to move or breathe, fueling panic that spiraled out of control. This catastrophe led to one of South Korea's worst stampede disasters, with 156 deaths and 170 crush injuries~\cite{sharma2023global}. 

These tragic incidents highlight the importance of crowd management at mass gathering events. The sheer number of people in a confined space can create a dangerous situation that can quickly spiral out of control, resulting in stampedes and crush injuries. Factors such as limited entry and exit points, uneven terrain, and unexpected choke points can exacerbate the risk of a stampede. Therefore, it is crucial for event organizers and authorities to implement effective crowd management strategies to prevent such incidents. This comprises actions such as appropriate scheduling of event timing, meticulous event venue planning and design, sufficient staffing, and unambiguous communication and signage. Effective crowd management not only ensures the safety and well-being of attendees but also helps to prevent damage to property and infrastructure. In light of the recent stampede incidents, it is clear that crowd management should be given the utmost importance in planning and executing mass gathering events. 

Crowd management is a multifaceted field that necessitates knowledge of engineering and technology, as well as comprehension of crowd behavior and crowd flow management, encompassing psychological and sociological aspects~\cite{sharma2018review}. By meticulous planning and execution, the objective of crowd management is to prevent crowd incidents~\cite{martella2017current}. Effective crowd management is a holistic process that includes several stages. It begins with meticulous planning before the event, considering all potential scenarios and preparing for them. During the event, the crowd needs to be closely monitored and controlled to ensure everyone's safety. After the event, it's important to gather feedback to understand what worked well and what didn't. Finally, these insights and lessons learned should be reported and used to improve crowd management strategies for future events. This approach ensures continuous improvement in managing crowds effectively~\cite{sharma2018review}.

In the pre-event planning stage, two primary technologies play a crucial role: crowd modeling and simulation, and social and web data mining. Crowd modeling and simulation enable the creation of virtual crowd scenarios, which serve as testing grounds for various crowd management strategies and their effectiveness. On the other hand, social and web data mining provides valuable insights into crowd demographics, behaviors, and trends. These insights help inform decision-making and enable the customization of crowd-management strategies to suit specific audience profiles. By leveraging these technologies in pre-event planning, crowd management can be approached with a greater level of knowledge, strategy, and effectiveness\cite{sharma2018review}.

As for the in-event control period, the acquisition of crowd data during monitoring, decision-making based on data analysis, and the implementation of crowd control measures are three key steps for success\cite{sharma2018review}. The primary goal of crowd control during the event is to detect instances of mass panic and respond quickly to dangerous situations. Various existing research proposed numerous methods for detecting crowd density to prevent surge incidents or to enforce social distancing. For instance, infrared thermal video sequences have been employed to monitor and estimate the density of crowds in real-time during large-scale public events~\cite{abuarafah2012real}. In addition, given the widespread Wi-Fi availability, it has been used to monitor crowd behavior and interaction~\cite{zhou2020understanding, weppner2016monitoring}.

Post-event feedback is crucial for preventing future incidents, and in this regard, social media data plays a pivotal role. The system for situational awareness can be enhanced by integrating feedback from the crowd and information related to the crisis that comes from social media. For example, in a system known as the HADRian, social media data was scrutinized after the Boston Marathon bombing in April 2013 to identify any unexploded or additional bombs\cite{ulicny2013situational}. Another example is Ushahidi, which is a versatile data collection, management, and visualization tool that enables data collection from multiple sources such as SMS, email, web, Twitter, and RSS, and offers robust features for post management and triaging through filters and workflows\cite{ushahidi}. Systems built on Ushahidi have been implemented worldwide in numerous situations, for instance, to oversee disaster relief efforts after the Haiti Earthquake in January 2010\cite{yuan2013harness}. 

Moreover, the outbreak of COVID-19 needed new real-time approaches for crowd monitoring and management systems for social distancing. Furthermore, Virtual Reality (VR) technology has been applied to analyze the emotional responses and stress levels of participants helps decision-makers gain enhanced insights into crowd management strategies for comparable occurrences~\cite{zhao2016crowd}.

%% file: 04_fairness_vadere.tex
\section{Fairness-aware Crowd Evacuation}\vspace{-1mm}
\label{Fairness section}

The importance of fairness in evacuating crowds lies in achieving a balanced distribution of evacuees across routes, ensuring equitable waiting times for different groups. Vulnerable groups, such as the elderly or pregnant women, are often overlooked in standard evacuation plans due to their physical limitations~\cite{UNwebsite}. The motivation for this section is to propose evacuation plans that ensure similar evacuation times for all individuals, regardless of their physical condition.

\subsection{Fairness Strategies}

Our experimental goal is to explore the design of evacuation routes for different groups of people in various crowd-gathering locations, aiming to achieve both high evacuation efficiency and fairness towards vulnerable populations. In real-life scenarios, the different running speeds of vulnerable groups and normal individuals can influence evacuation times, potentially leading to hazards like pushing or tripping. To address this, we propose a strategy where a dedicated evacuation exit is designated exclusively for vulnerable groups, guided by mobile notifications or other means, while other individuals can use the nearest exit. We hypothesize that this design can reduce overall evacuation time, especially for vulnerable groups, ensuring efficiency and fairness simultaneously. To validate our hypothesis, we proposed three strategies across multiple crowded event scenarios to assign the evacuation gate:
\begin{itemize}[noitemsep, leftmargin=*]
\item \textbf{Randomly gate assignment (RGA):} Individuals evacuate by randomly selecting a gate without any specific guidance.
\item \textbf{Vulnerable people exclusive gate assignment(VEGA):} Vulnerable individuals are directed to a designated gate exclusively. Healthy people are assigned the closest gate.
\item \textbf{Closest gate assignment (CGA):} All individuals are assigned the closest gate regardless of their physical state.
\end{itemize}

\subsection{Fairness Metric: Normalized Evacuation Time Disparity (NETD) }
We compare these strategies by introducing a fairness metric; Normalized Evacuation Time Disparity (NETD). 
The Normalized Evacuation Time Disparity (NETD) metric can be defined as the difference in average evacuation times between vulnerable and healthy people, normalized by the overall average evacuation time for the entire population. This metric captures the relative disparity in evacuation efficiency between the two groups, providing a standardized measure of fairness.

{\small
\begin{align}\label{eq:NETD}
NETD = \frac{| \text{Avg. Time}_{vul} - \text{Avg. Time}_{heal} |}{\text{Avg. Time}_{all}}
\end{align}
}

$\text{Avg. Time}_{vul}$ is the average evacuation time for vulnerable people, while $\text{Avg. Time}_{heal}$ is the average evacuation time for healthy people. $\text{Avg. Time}_{all}$ is the weighted average evacuation time for the entire population, calculated as:

{\small
\begin{align}
&\text{Avg. Time}_{all} = \frac{N_{vul} \times \text{Avg. Time}_{vul} + N_{heal} \times \text{Avg. Time}_{heal}}{N_{vul} + N_{heal}}
\end{align}
}

$N_{vul}$ is the number of vulnerable people, while $N_{heal}$ is the number of healthy people.

This can be interpreted as an NETD of $0$ indicating perfect fairness, with identical average evacuation times for vulnerable and healthy people, adjusted for population proportions. A higher NETD value indicates a greater disparity in evacuation times, signaling less fairness.

The NETD model aligns with Rawlsian principles in that it aims to measure and address inequalities between different groups, specifically vulnerable and healthy populations, in evacuation scenarios~\cite{nagel1973rawls,Fleurbaey_1995}. By focusing on the disparity in evacuation times and aiming to minimize this disparity, the model encapsulates a core aspect of the Rawlsian Difference Principle: making inequalities work to benefit the least advantaged, in this case, the vulnerable population. To relate this directly to Rawlsian philosophy, the NETD metric is a proxy for assessing the ``fairness'' of an evacuation strategy by quantifying disparities between different groups. This is conceptually consistent with Rawls's focus on societal structures that protect and benefit the least well-off. A Rawlsian approach would prioritize reducing the NETD value towards zero, symbolizing that the vulnerability does not lead to a disadvantage in evacuation times\footnote{We provide a philosophical take on Rawls's principles in the context of evacuation systems in the Appendix.}. 

\vspace{-1mm}
\subsection{Modeling and Simulation}\vspace{-1mm}
We used Vadere~\cite{kleinmeier2019vadere} to simulate various crowd event setups which we call a map. Four evacuation exits were placed at the corners of the map. 
\paragraph{Crowd Statistics} On the map, there are $1363$ people, consisting of $340$ vulnerable people and $1023$ healthy normal people which counts for $25\%$ of the crowd\footnote{Based on the National Environmental Public Health Tracking from Centers of Disease Control and Prevention (CDC), the populations with vulnerabilities and health status identified as a disability among adults $>=18$ years of age is around $25\%$ in the United States in 2021~\cite{CDChealthtrack}.}. 

\paragraph{Human behavior} The average running speed of healthy and young individuals (aged 20-45 years) is $\approx 5.4$ miles per hour or $\approx 2.4$ meters per second~\cite{lung2021effects}. Hence, we set the average speed of normal people to be $1.0-1.3$ meters per timestep. Each timestep represents $0.48$ seconds. Vulnerable people, such as the elderly, move at a slower pace as the speed of humans decreases by $20\%$ every decade~\cite{devita2016relationships}. Therefore, we set the vulnerable people's speed to be half that of normal people, which is $0.5-0.6$ meters per timestep \footnote{We assume various groups of vulnerable people have similarly limited mobility. We grouped children, the elderly, and the disabled under the singular category of ``vulnerable people'' and assigned them the same movement speed.} .

\paragraph{Crowd behavior} We exploited the Optimal Steps Model (OSM) in Vadere which incorporates the psychological principle of ``social distance'' into its mathematical framework, which means that crowd strive to avoid encroaching on others' personal or intimate space and to prevent physical contact~\cite{kleinmeier2019vadere}.

\begin{figure}[!t]
\centering

\begin{minipage}{1\linewidth}
    \makebox[.25\linewidth]{\includegraphics[width=.2\linewidth]{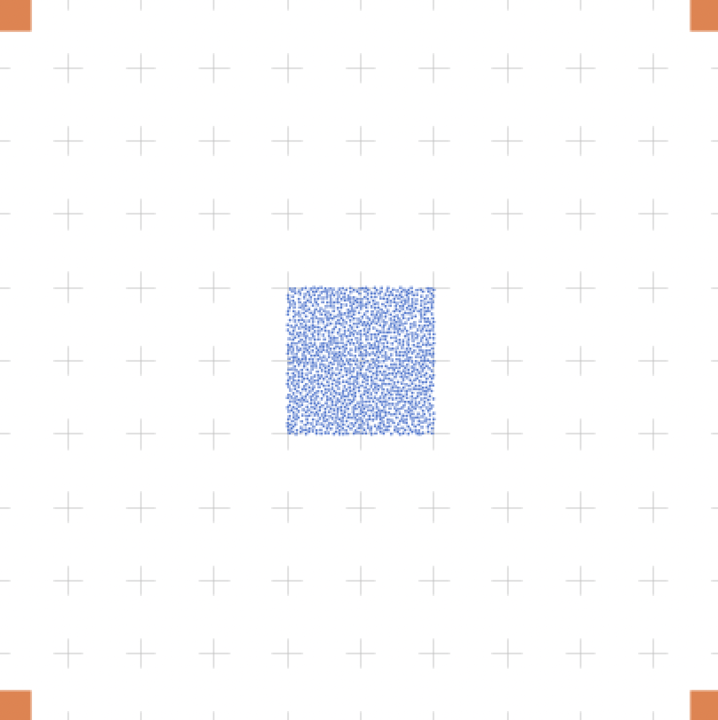}}%
    \makebox[.25\linewidth]{\includegraphics[width=.2\linewidth]{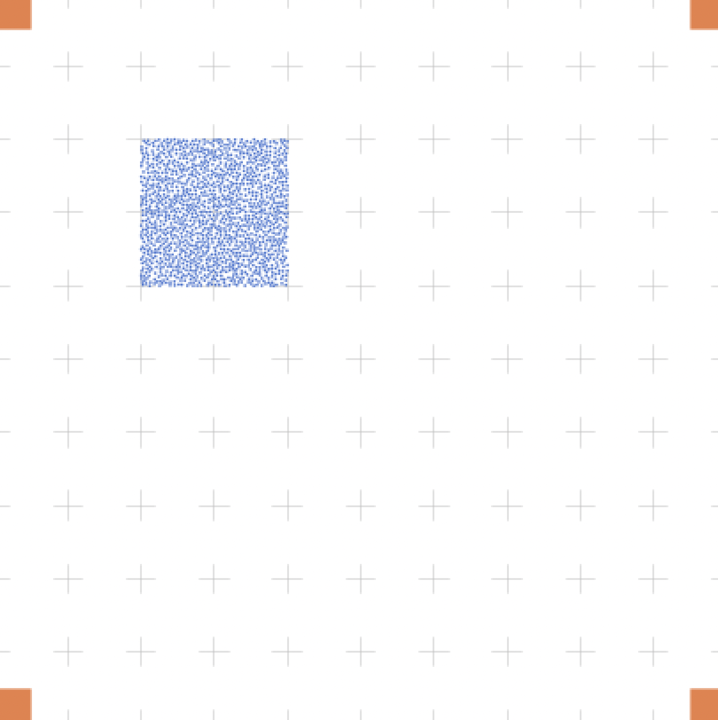}}%
    \makebox[.25\linewidth]{\includegraphics[width=.2\linewidth]{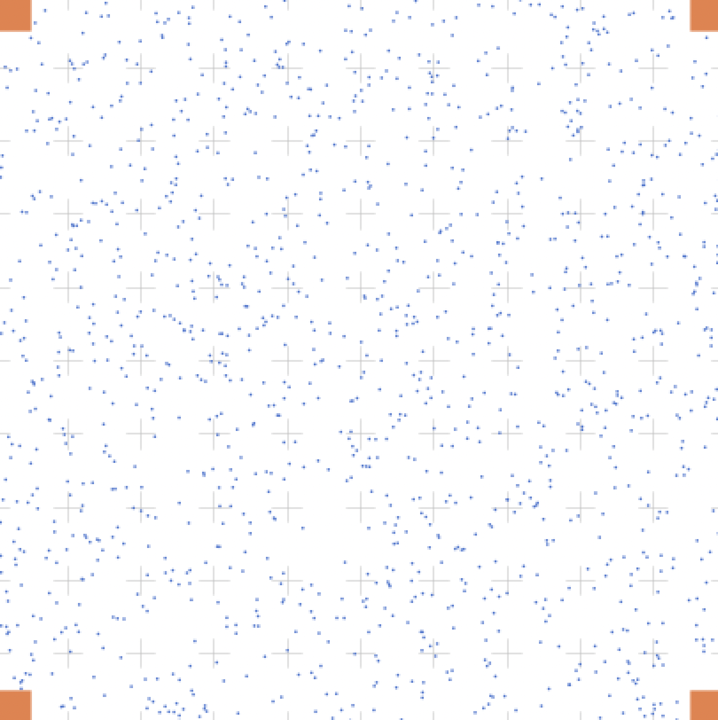}}%
    \makebox[.25\linewidth]{\includegraphics[width=.2\linewidth]{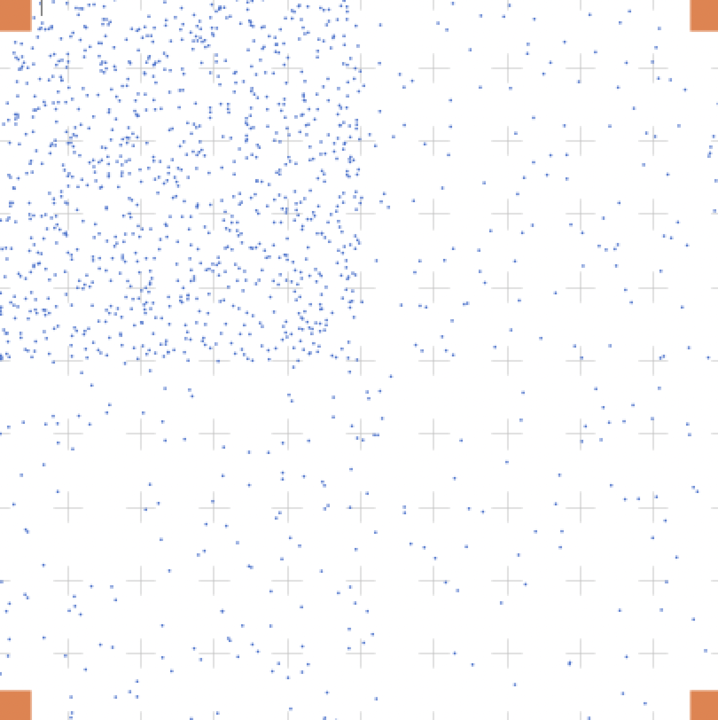}}
    \makebox[.25\linewidth]{\small (a)}%
    \makebox[.25\linewidth]{\small (b)}%
    \makebox[.25\linewidth]{\small (c)}%
    \makebox[.25\linewidth]{\small (d)}
\end{minipage}%

\caption{\small{The four scenarios of crowd distribution in Vadere, where blue dots represent individuals, and orange blocks represent exit locations. (a): Center crowd gathering; (b): Non-center crowd gathering; (c): Evenly crowd dispersing; (d): Unevenly crowd dispersing} \vspace{-3mm}
\label{fig:vadere_example}}
\vspace{-3mm}
\end{figure}

\vspace{-1mm}
\subsection{Evaluation}\vspace{-1mm}
We simulated various scenarios for crowd distribution on the map illustrated in Figure~\ref{fig:vadere_example}:

\begin{itemize}[noitemsep, leftmargin=*]
\item \textbf{Scenario 1: Center crowd gathering:} Serving as an exemplar for an event with a setup focused on the center stage.
\item \textbf{Scenario 2: Non-center crowd gathering:} Serving as an exemplar for an event with a setup focused on a non-center stage.
\item \textbf{Scenario 3: Evenly crowd dispersing:} Representing carnival event across the entire map.
\item \textbf{Scenario 4: Unevenly crowd dispersing:} Representing carnival event with varying area of crowd densities: $75\%$ in top-left, $10\%$ in top-right, $10\%$ in bottom-left, and $5\%$ in bottom-right.
\end{itemize}

We measure the fairness using the fairness metric (NETD) described in Equation~\ref{eq:NETD}. 
Table~\ref{tbl:fairness} in the Appendix shows the detailed results of average evacuation time and NETD using three different strategies. Gate Time $(G_1, G_2, G_3, G_4)$ shows each gate's last human exiting time, providing insight into the gate utility. In summary, RGA shows the worst evacuation time for $all$ population in all scenarios while CGA demonstrates the best average evacuation time for $all$ population. 
In scenarios $S_2$ and $S_4$, VEGA achieves the lowest NETD (better fairness) of $0.16$ and $0.49$, respectively, with a slower average evacuation time for $all$ population compared to CGA. \textbf{This result highlights that optimizing for the average evacuation time for all population can undermine the vulnerable populations.} In contrast, VEGA achieves the highest NETD (worst fairness) in scenario $S_3$ with a value of $1.76$. 

We also evaluated the gate utility ($G$), indicating how long a particular gate was used for evacuation. To compare the gate utility, we look into which strategy can achieve a similar utility across all gates. The insight is to ensure all gates are used and not a particular one with a crowd surge. Hence, we compute the Euclidean distance for each vector of the gate times $(G_1, G_2, G_3, G_4)$.  Table~\ref{tbl:gatetimes} in the Appendix shows the detailed results. In summary, since VEGA has a dedicated gate for vulnerable people, this particular gate time ($G_1$)is higher in all the scenarios. However, CGA showed better similarity in gate times except for $S_2$ where the population is not centered (Figure~\ref{fig:vadere_example}).

\paragraph{\textbf{Takeaways}} Our preliminary experiment aims to investigate whether a specific evacuation strategy for vulnerable groups enhances fairness while maintaining overall efficiency (gate utility). In crowd scenarios, $S_1$ and $S_3$, a vulnerable exclusive gate may not improve fairness or efficiency when people initially gather around the center of each exit. Nonetheless, when many vulnerable individuals congregate near a single exit, such as in scenarios $S_2$ and $S_4$, the VEGA strategy leads to a fairness improvement of $78\%$ on average, compared to RGA and CGA.
In scenario $S_2$, the VEGA strategy demonstrates a fairness increase of $76.81\%$ and $77.46\%$ compared to RGA and CGA, while in scenario $S_4$, it shows a fairness improvement of $80.2\%$ and $78\%$ compared to RGA and CGA. Hence, the average VEGA improvement from RGA and CGA for $S_2$ and $S_4$ is $78.135\%$.

On the other hand, CGA shows a better fairness value (NETD) for scenarios $S_1$ and $S_3$. In particular, when the crowd is centered ($S_1$) or evenly dispersed ($S_3$), CGA can show a slight fairness increase of $1.4\%$ and $2.6\%$, respectively, compared to RGA. However, compared to VEGA, CGA can improve fairness for $S_1$ and $S_3$ by $24.5\%$ and $57.4\%$ respectively. Hence, the average CGA improvement from RGA and VEGA for $S_1$ and $S_3$ is $21.5\%$. From this preliminary analysis, we conclude that one strategy can not fit all crowd scenarios, and the evacuation strategy has to be adaptive based on the current state of the crowd\footnote{In real-world scenarios, it is anticipated that some individuals may not adhere to the recommended guidelines set by the organizers or the automated crowd management system. Therefore, a comprehensive study of the crowd's social and psychological dynamics is imperative to better understand and address such situations.}.

%% file: 05_preventive_netlogo.tex
\vspace{-1mm}
\section{Preventative Strategy}
\vspace{-1mm}

We recognize the pivotal role of preventive measures in mitigating harm, underscoring their significance compared to reactive evacuation plans post-crowd surge incidents.

Our second experiment is spurred by the tragic Astroworld Festival accident in 2021~\cite{astroworld}. This festival featured a main stage for the primary performance and a secondary stage hosting performances by other artists throughout the day. During the concert night, following a performance at the secondary stage, the audience gravitated towards the already congested area near the main stage, resulting in a surge and crush. The repercussions were dire, with multiple fatalities and areas densely packed to the extent of providing only $1.85$ square feet per person~\cite{astroworld}. A visual illustration is shown in Figure~\ref{fig:astroworldillustration} in the Appendix.

To tackle congestion during stage performances, our prevention strategy advocates for stage switching at designated intervals. This approach aims to mitigate overcrowding and enhance crowd management throughout the event. To determine optimal switching points between stages, we introduce three metrics: the Panic State and Surge State per individual, along with the Crowded State for each subarea. Further elaboration on these metrics will be provided in Section~\ref{section4modeling}.

To achieve these objectives, we have developed a simulation tool using NetLogo, an agent-based modeling environment~\cite{tisue2004netlogo, zia2020agent}, with customizable attributes such as position and walking speed.

\vspace{-1mm}
\subsection{Modeling and Simulation}\vspace{-1mm}
\label{section4modeling}

We create a simulation of a crowded environment featuring two stages within a two-dimensional square world consisting of $51\times51$ patches. Each patch is represented by an xy-coordinate point, with the origin located at $(0, 0)$ in the bottom left corner. The map is subdivided into multiple subareas using a $5\times5$ grid, with each subarea containing 25 patches. Individuals, depicted as triangular shapes, are situated on these patches and utilize information from their patch as well as neighboring patches and individuals to inform their decision-making processes. The simulation environment includes two stages, a bar, and a restroom, positioned at the far left and right sides for the stages and the top and bottom sides for the bar and restroom, respectively. A visual representation of the map is provided in Figure~\ref{fig:netlogoexample} in the Appendix.

To model the human behavior in the simulation, we consider four factors in our tool:
\begin{itemize}[leftmargin=*, topsep=0pt]

\item \textbf{Speed Variation:} Half of the individuals are randomly assigned a speed of 1 step per time step, while the other half are set to 2 steps per time step using the ``random'' method in NetLogo's code.
\item \textbf{Comfort Zone and Preferred Distance:} Acknowledging individual preferences, we randomly assign a comfort distance between 1-10 meters to each agent in the NetLogo code, reflecting varying levels of comfort regarding proximity to the stage.
\item \textbf{Bar and Restroom Visits:} Individuals randomly visit the bar or restroom during the simulation. The NetLogo interface allows users to set the total time spent and the frequency of these visits. We set by default 40\% of individuals visit the bar or restroom every $50$ time-steps, with each trip lasting $50$ time-steps. Additionally, we provide adjustment bars in the interface to customize the frequency and duration of these visits if needed in the simulation.
\item \textbf{Hesitation Time for Stage Switching:} 
This captures the variability in the time individuals take to decide whether to switch stages after a performance has concluded. We assign a random hesitation time between $1-20$ time-steps to each individual. 

\end{itemize}

\subsection{Prevention}
Our prevention strategy focuses on determining the best time to switch stages, considering various parameters. To achieve this, we have developed a simulation tool to evaluate different scenarios and parameters, allowing us to estimate the most suitable moment for switching the performance to a new stage. 
We propose using a set of metrics to determine the status of individuals and subareas during the event:

\begin{itemize}[noitemsep, leftmargin=*, topsep=0pt]
    \item \textbf{Panic state:} Studies in crowd psychology indicate that individuals experiencing confinement or limitations in their mobility may develop sensations of panic or anxiety~\cite{helbing2012crowd}. Consequently, our simulations posited that an individual transitions into a state of panic when they encounter obstructions or delays en route to facilities such as restrooms or bars, persisting for a period surpassing a predefined panic threshold (PT).
    \item \textbf{Surge state:}  A ``surge state'' in crowd management can be inferred as a condition or situation where there is a sudden and significant increase in crowd density that exceeds normal potentially leading to congestion, restricted movement. Our simulations suggest that an individual is considered to be in a surge state when they find themselves obstructed and unable to proceed towards their destination, such as a stage, for a period longer than a predetermined surge threshold (ST).  
    \item \textbf{Crowded state:} A subarea enters the crowded state when over $70\%$ of its patches are occupied by people, and at least one person within the subarea is in either panic or surge state.
   \item \textbf{Switch index (SI):} 
   \begin{itemize}[noitemsep,leftmargin=*, topsep=0pt]
   \item The switch index (SI) represents a threshold on the duration for which a subarea remains continuously in the crowded state before the performance is switched to the other stage.
    \item When a subarea consistently stays in a crowded state, surpassing the threshold of the switch index (SI), and its two neighboring subareas are also in a crowded state (though not necessarily exceeding the SI threshold), it indicates a critical surge situation.
    \item At this point, the currently performing stage receives instructions to halt, and the performance switches to another stage.
   \end{itemize}   
\end{itemize}

\begin{figure*}[!t]
\centering
\begin{minipage}{1\linewidth}
\centering
    \makebox[.2\linewidth]{\includegraphics[width=.2\linewidth]{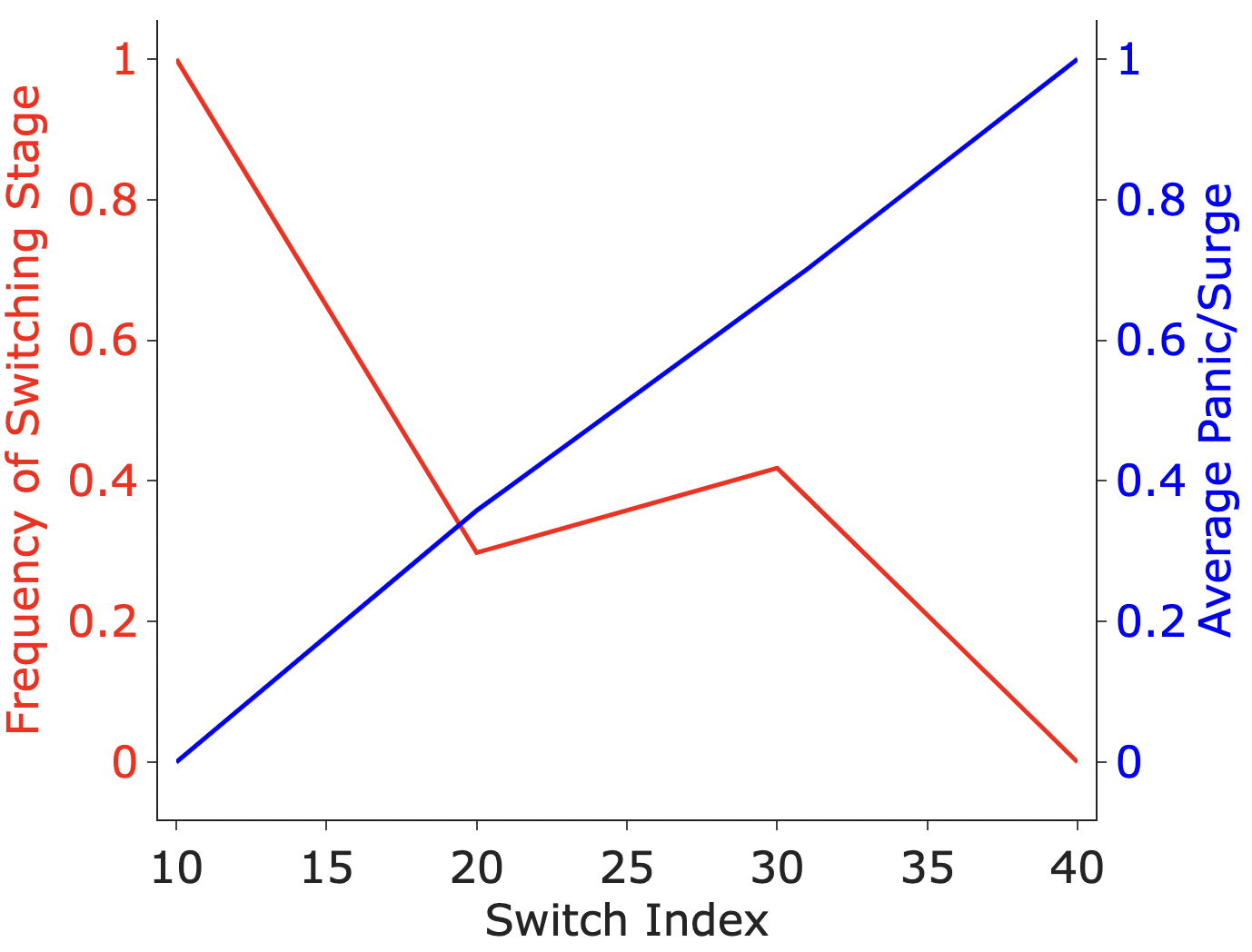}}%
    \makebox[.2\linewidth]{\includegraphics[width=.2\linewidth]{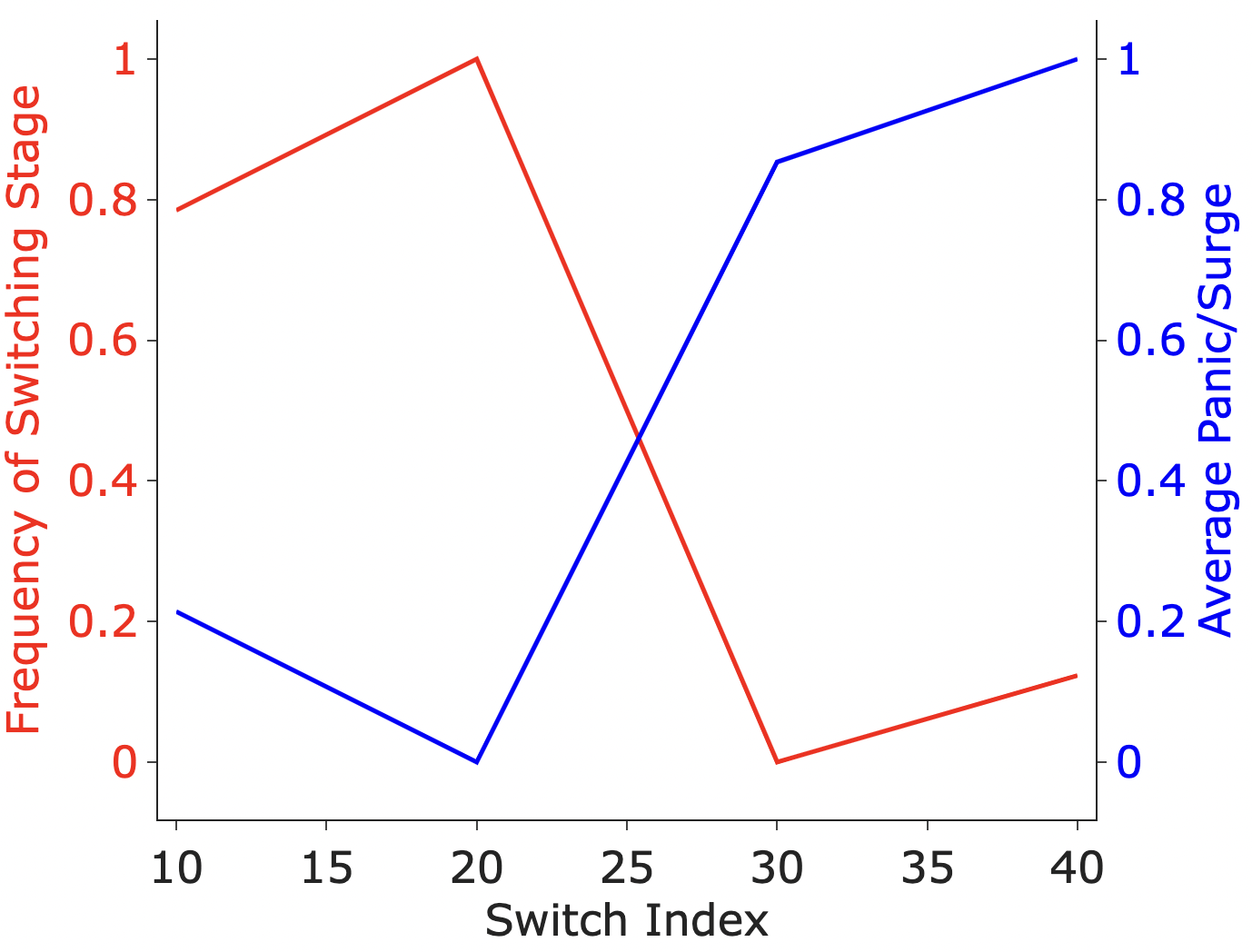}}%
    \makebox[.2\linewidth]{\includegraphics[width=.2\linewidth]{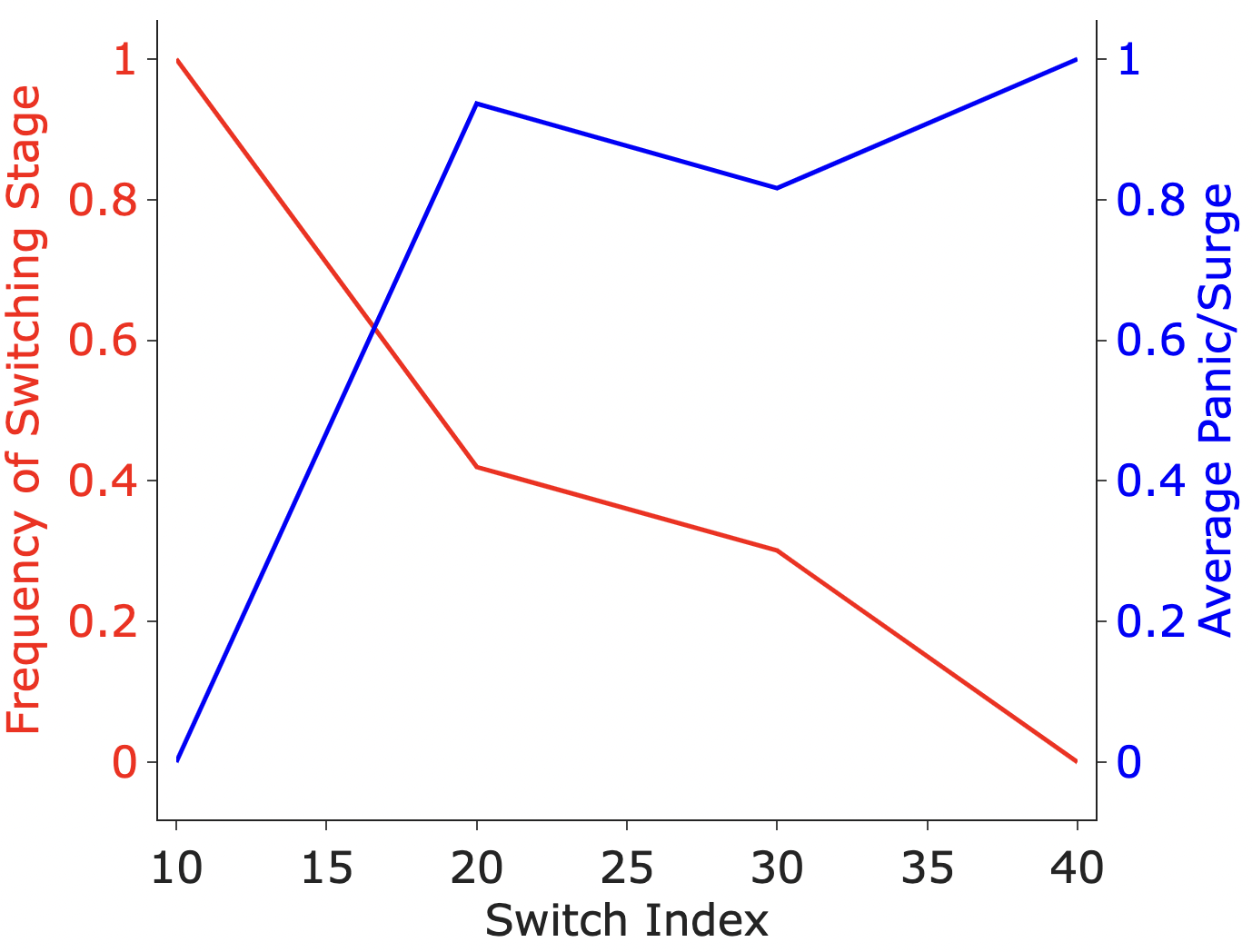}}%
    \makebox[.2\linewidth]{\includegraphics[width=.2\linewidth]{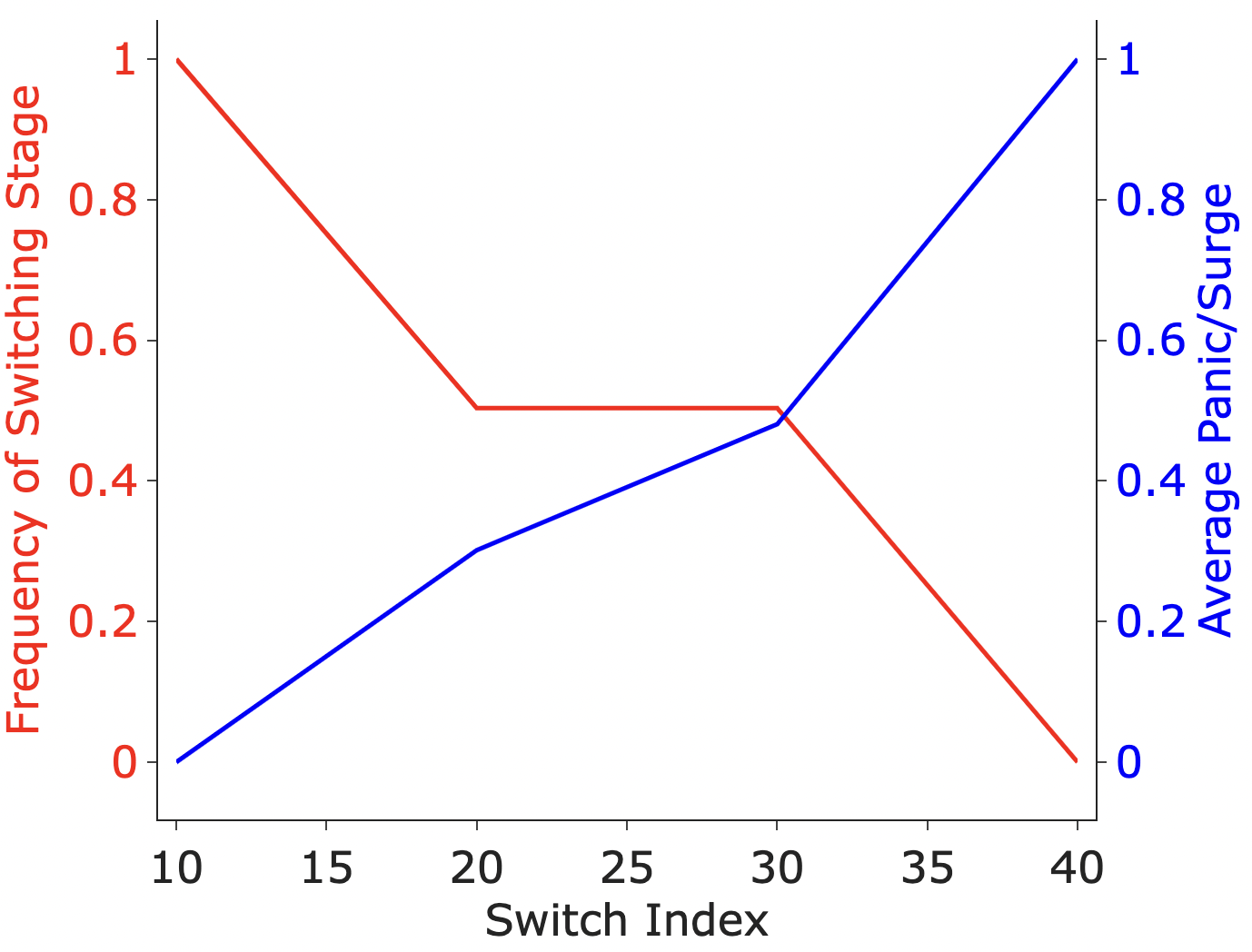}}%
    \makebox[.2\linewidth]{\includegraphics[width=.2\linewidth]{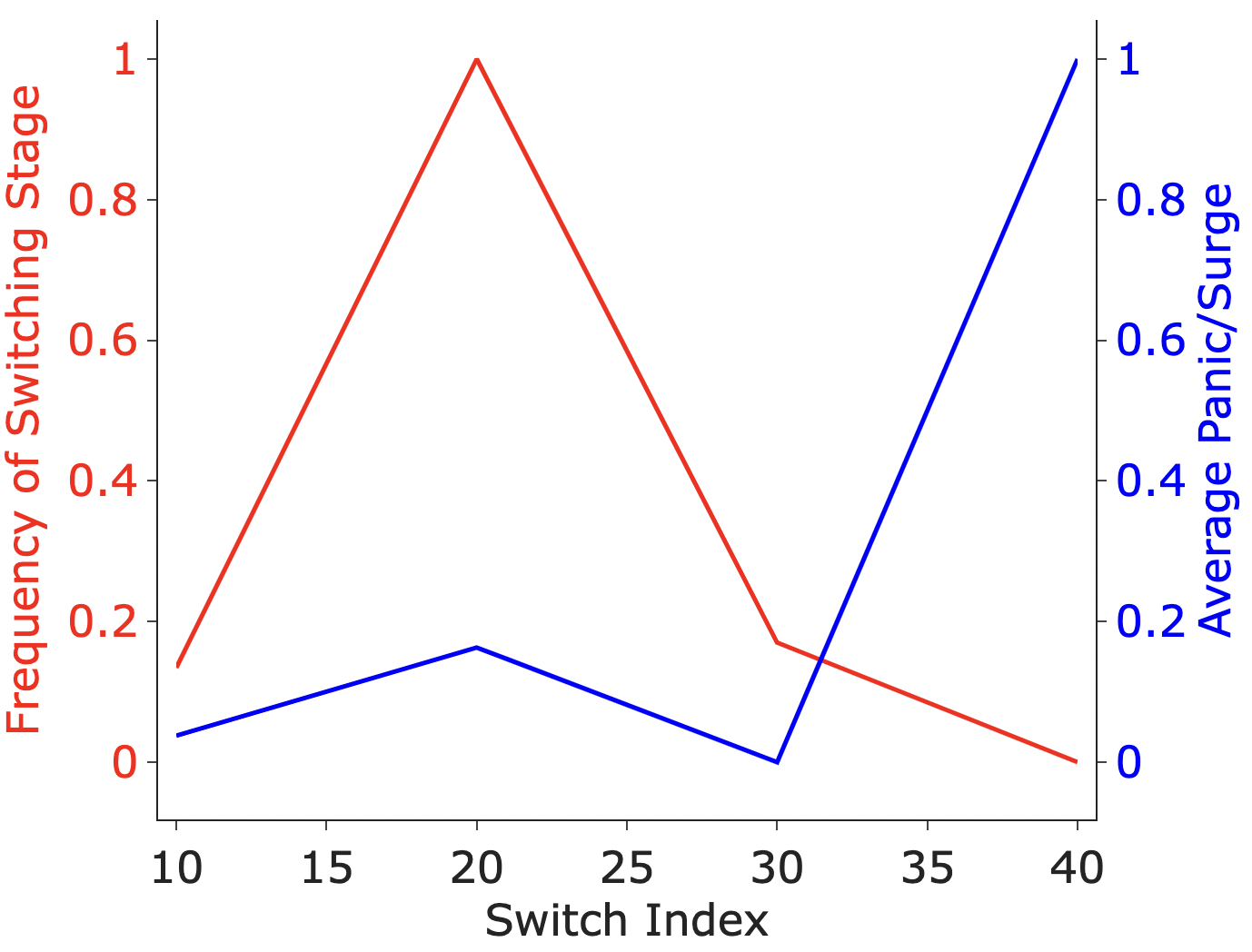}}%
    
    \makebox[.2\linewidth]{\small (a)}%
    \makebox[.2\linewidth]{\small (b)}%
    \makebox[.2\linewidth]{\small (c)}%
    \makebox[.2\linewidth]{\small (d)}%
    \makebox[.2\linewidth]{\small (e)}%

\end{minipage} 

\caption{\small{Frequency of switching performance to another stage (F) and Average Panic/Surge value with four different Switch Index (SI) (10, 20, 30, 40) using different parameters. (a): PN500, BRF50, PT10, ST30; (b): PN750, BRF50, PT10, ST30; (c): PN500, BRF30, PT10, ST30; (d): PN500, BRF50, PT10, ST40; (e): PN500, BRF50, PT20, ST30.}}\vspace{-5mm}
\label{fig:5results}

\end{figure*}

\vspace{-1mm}
\subsection{Evaluation} \vspace{-1mm}

We start our evaluation by investigating the correlation between stage positions and the probability of surge accidents. The tragic incident at Astroworld underscores the profound impact of stage proximity on crowd dynamics. We simulate three maps for stage positions scenarios. Map A features stages directly facing each other, Map B positions one stage in the bottom-right corner and another in the middle on the left side, while Map C situates two stages at the top-left and bottom-right, respectively. Figure~\ref{fig:3stages} in the appendix presents a visual representation to these three distinct scenarios. In each map, half of the individuals move at twice the speed of the other half. Their comfort distances are randomly assigned between $1-10$ units patch-size, and hesitation times are randomly distributed between $1-20$ time-steps. The Switch Index (SI) is set at $10$ time-steps. Additionally, $40\%$ of individuals are designated to go to the bar or restroom, with each trip lasting for $50$ time-steps. Furthermore, we have identified four adjustable parameters: total number of people (PN), frequency of bar/restroom visits (BRF), panic threshold (PT), and surge threshold (ST). The default parameter settings are as follows: PN $= 500$, BRF $= 50$, PT $= 10$, and ST $= 30$. We employ several metrics to evaluate crowd behavior, namely the frequency of stage switching (F) and Average Panic/Surge (APS). The Average Panic/Surge (APS) metric denotes the average number of individuals in Panic/Surge Status per time-step. These metrics collectively enable us to assess the efficacy of our strategies and response measures. More details on these values for each Map scenario is shown in Figure~\ref{fig:3stages} in the Appendix. On average, Map C demonstrates a reduction in F by $26\%$ and APS by $34\%$. These results underscore the effectiveness of increasing the distance between stages in mitigating the likelihood of crowd surge incidents and reducing the frequency of stage performance switching. Continuing the evaluation, subsequent experiments will focus on Map C to investigate the impact of varying values of SI on F and APS across different parameters, including:

\begin{itemize}[noitemsep, leftmargin=*, topsep=0pt]
\item \textbf{Total number of people (PN):} Initially set at 500 in Figure~\ref{fig:5results}(a), PN increases to 750 in Figure~\ref{fig:5results}(b), portraying a denser environment.
\item \textbf{Frequency of bar/restroom use (BRF):} Initially set at 50 in Figure~\ref{fig:5results}(a), BRF decreases to 30 in Figure~\ref{fig:5results}(c), indicating restricted access to restroom and bar facilities. 
\item \textbf{Panic threshold (PT) and Surge threshold (ST):} Different values of ST and PT represent diverse audience compositions. Figure~\ref{fig:5results}(a) features ST=10 and PT=10. In Figures~\ref{fig:5results}(d) and (e), ST increases to 40, and PT increases to 20, respectively. 
\end{itemize}

\paragraph{\textbf{Takeaways}}

An increase in the total number of people (PN) increases the likelihood of crowded states, potentially raising both the frequency of stage switching (F) and the APS. However, the impact on F and APS may vary depending on the SI chosen. For example, in Figure~\ref{fig:5results}(b), the denser crowd (PN750) might require a different SI to manage the increased density without causing panic or surge states.
Decreasing the frequency of bar/restroom use (BRF) to 30, as shown in Figure~\ref{fig:5results}(c), could indicate fewer exits and entries to these facilities, which may reduce movement and potential blockage. However, if access is too restricted, it might increase the APS due to individuals being blocked for longer periods, exceeding the PT.
Various combinations of PT and ST values indicate different crowd behaviors. For example, a higher PT (indicating a higher tolerance for being blocked) may lead to a lower APS, as individuals would not enter a panic state as quickly. However, if the ST is also high, individuals may block the way to the stage for longer, potentially increasing the crowded state and APS, as seen in Figures~\ref{fig:5results}(d) and (e).
The SI's role is crucial in managing the balance between the frequency of stage switches and APS. A low SI may lead to frequent disruptions (high F) but could keep APS low by preventing sustained crowded states. Conversely, a high SI might lead to fewer disruptions but could allow crowded states to persist longer, potentially increasing APS.

The conclusions that can be drawn from these figures, with the given metrics, underscore the delicate balance needed in crowd management. An effective SI value would minimize the crowded state's duration, thus reducing the average panic/surge while also optimizing the frequency of stage switches to maintain a smooth event flow. 

\paragraph{\textbf{Beyond simulation}} The estimation of individual states (panic/surge state) can be facilitated by smartphones and wearable sensors, including accelerometers, gyroscopes, and heart rate monitors. These sensors enable the assessment of gait parameters, thereby providing a reliable method for determining individuals' states.  
Besides, indoor GPS and Wi-Fi can be utilized to infer locations. The system could require individuals to input essential personal data for identification and support during crises. It would then use this data to generate evacuation routes through a user-friendly app interface, alerting users of deviations from the recommended path. This strategy is based on the belief that people are more likely to follow evacuation suggestions if they perceive the procedures as fair and just~\cite{10.1111/j.1745-9125.2012.00289.x}.

%% file: 06_conclusion.tex
\vspace{-2mm}
\section{Future Work \& Conclusion}\vspace{-1mm}

As urbanization and population density continue to rise, ensuring efficient and safe evacuation procedures during emergencies or crowded events becomes increasingly challenging.
Our research focuses on addressing crowd management challenges by exploring both evacuation strategies and preventive methodologies. We emphasize the importance of balancing fairness and efficiency in evacuation plans while considering psychological factors influencing individual social distancing behavior during evacuations. Our preliminary results showed that by utilizing simulation tools like Vadere, we can design evacuation strategies that consider vulnerable people. Additionally, we developed a NetLogo-based tool to simulate preventive strategies based on the crowd current state. 
In the future, we aim to further optimize our fairness-aware evacuation and preventive approach by integrating different social dynamics including couples, families, and friends, who have a tendency to select the same escape route during an evacuation, populations who will not abide by the suggested routes of evacuation, post-event analysis, and privacy protection measures. These steps will ensure ethical data usage and enhance the overall effectiveness of crowd management.

%% file: 07_appendix.tex
\onecolumn
\section*{Appendix}

\subsection*{Rawlsian principles in the context of evacuation systems }
The NETD model in our fairness strategy is built with Rawlsian principles. At a high level, Rawlsian philosophy advocates for a societal structure where decisions are made under a veil of ignorance. The veil of ignorance posits that individuals, when unaware of their own position in society, would agree to fair terms that benefit all, including vulnerable populations~\cite{10.1073/pnas.1910125116}. In the context of evacuation systems, this principle is applied through mechanisms where individuals provide basic information without knowledge of how the vulnerability is defined or measured in that specific scenario. This uncertainty about whether one might be classified as vulnerable effectively motivates collective protection of the vulnerable, which enhances the probability of one's adherence to recommended evacuation routes. In doing so, the system ensures that emergency evacuations are executed with a heightened sense of communal responsibility and solidarity.

\begin{figure}[ht]
\centering
\includegraphics[scale=0.5]{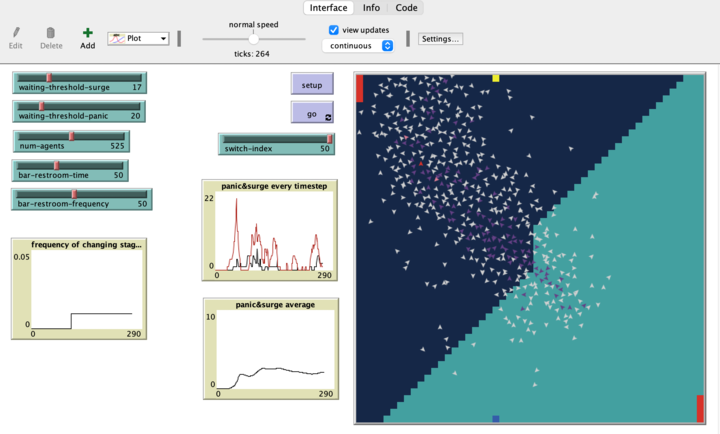} 
\caption{Our proposed simulator tool interface utilizing NetLogo. The top-left and bottom-right red rectangles on the simulation map denote two stages. Additionally, the upper yellow dot and lower blue dot symbolize the restroom and bar, respectively. The blue area indicates patches proximal to the left stage, while the green area depicts patches adjacent to the right stage. 
}\vspace{-3mm}
\label{fig:netlogoexample}
\end{figure}

\begin{table*}[ht]
  \caption{Comparison of average evacuation times for vulnerable ($V$) and healthy people ($H$) and average across all population ($all$), fairness metric $NETD$, and Gate Time $(G1, G2, G3, G4)$ using $3$ different strategies under $4$ different scenarios ($S_1, \dots, S_4$). Time is measured in the simulation step. 
  \label{tbl:fairness}}
    \centering  
  \begin{tabularx}{0.68\textwidth}{|l|l|l|l|l|l|l|l|l|l|}
    \hline
    \multirow{2}{*}{} &
        \multicolumn{3}{c|}{RGA} & \multicolumn{3}{c|}{VEGA} & \multicolumn{3}{c|}{CGA} \\
    \cline{2-10}
    & \multicolumn{2}{c|}{$(V, H, all)$} & NETD &  \multicolumn{2}{c|}{$(V, H, all)$} & NETD &  \multicolumn{2}{c|}{$(V, H, all)$} & NETD  \\
    \hline
    $S_1$ & \multicolumn{2}{c|}{(149.9, 80.9, 98.1)} & 0.70 & \multicolumn{2}{c|}{(154.3, 72.4, 93)} & 0.88 & \multicolumn{2}{c|}{(125.7, 68.3, \textbf{82.8})} & \textbf{0.69}  \\
    \hline
     $S_2$ & \multicolumn{2}{c|}{(153.4, 83.8, 101.1)} & 0.69  & \multicolumn{2}{c|}{(97.0, 83.1, 86.76)} & \textbf{0.16} & \multicolumn{2}{c|}{(132.1, 70.9, \textbf{86.4})} & 0.71  \\
     \hline
    $S_3$ & \multicolumn{2}{c|}{(138.0, 70.5, 87.5)} & 0.77  & \multicolumn{2}{c|}{(137.4, 33.0, 59.17)} & 1.76  & \multicolumn{2}{c|}{(64.7, 33.5, \textbf{41.4})} & \textbf{0.75}  \\
    \hline
    $S_4$ & \multicolumn{2}{c|}{(140.7, 69.5, 87.5)} & 0.81 & \multicolumn{2}{c|}{(91.2, 58.3, 66.65)} & \textbf{0.49}  & \multicolumn{2}{c|}{(79.0, 41.8, \textbf{51.2})} & 0.73  \\\hline
\end{tabularx}

\end{table*}

\begin{table*}[ht]
  \caption{Comparison of Gate Time $(G1, G2, G3, G4)$ using $3$ different strategies under $4$ different scenarios ($S_1, \dots, S_4$). Euclidean distance (Euc.D) is used to measure how close the gate times across the 4 gates. Time is measured in the simulation step. 
  \label{tbl:gatetimes}}
    \centering  
  \begin{tabularx}{0.71\textwidth}{|l|l|l|l|l|l|l|}
    \hline
    \multirow{2}{*}{} &
        \multicolumn{2}{c|}{RGA} & \multicolumn{2}{c|}{VEGA} & \multicolumn{2}{c|}{CGA} \\
    \cline{2-7}
    & $(G_1, G_2, G_3, G_4)$ & Euc.D &  $(G_1, G_2, G_3, G_4)$ &  Euc. D & $(G_1, G_2, G_3, G_4)$ & Euc. D\\
    \hline
    $S_1$  & (201, 187, 195, 195) & 19.90 & (198, 99, 88, 98)&  179.16 &  (143, 144, 146, 150) & \textbf{10.72}\\
    \hline
     $S_2$ & (143, 193, 250, 200) & 151.64 & (140, 105, 115, 103) & \textbf{58.87} & (90, 153, 204, 158) &  156.39\\
     \hline
    $S_3$ & (263, 254, 227, 257) & 55.24 & (262, 61, 63, 63) & 345.83 &  (122, 116, 129, 112)& \textbf{25.66}\\
    \hline
    $S_4$ & (226, 233, 258, 250) & 51.25  &  (260, 93, 60, 90) & 314.38 &  (132, 122, 120, 116) &  \textbf{23.59}\\\hline
\end{tabularx}

\end{table*}

\begin{figure}[ht]
\centering
\begin{minipage}{.5\textwidth}
    \centering
    \includegraphics[scale=0.6]{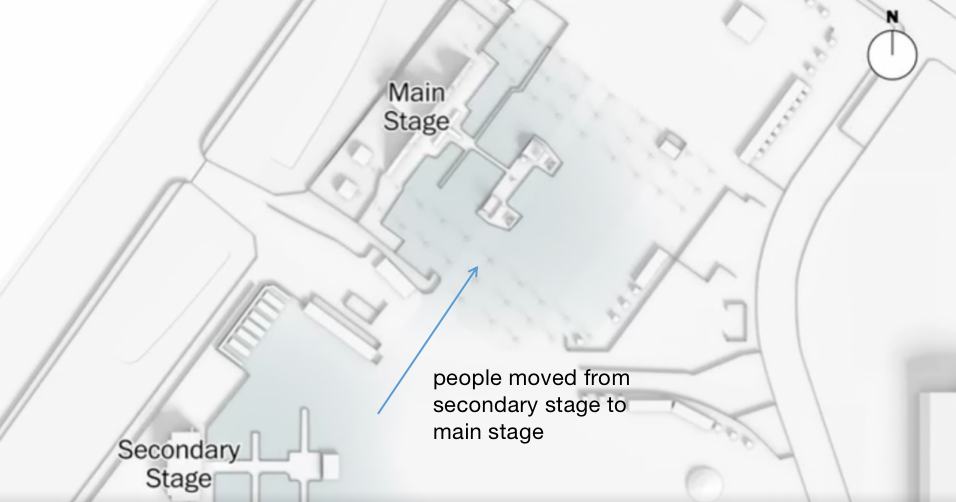} 
    \caption{The illustration of the 2021 Astroworld Festival accident.}\vspace{-3mm}
    \label{fig:astroworldillustration}
\end{minipage}%
\end{figure}

\begin{figure*}[ht]
     \centering
     \begin{subfigure}[b]{0.3\textwidth}
         \centering
         \includegraphics[width=\textwidth]{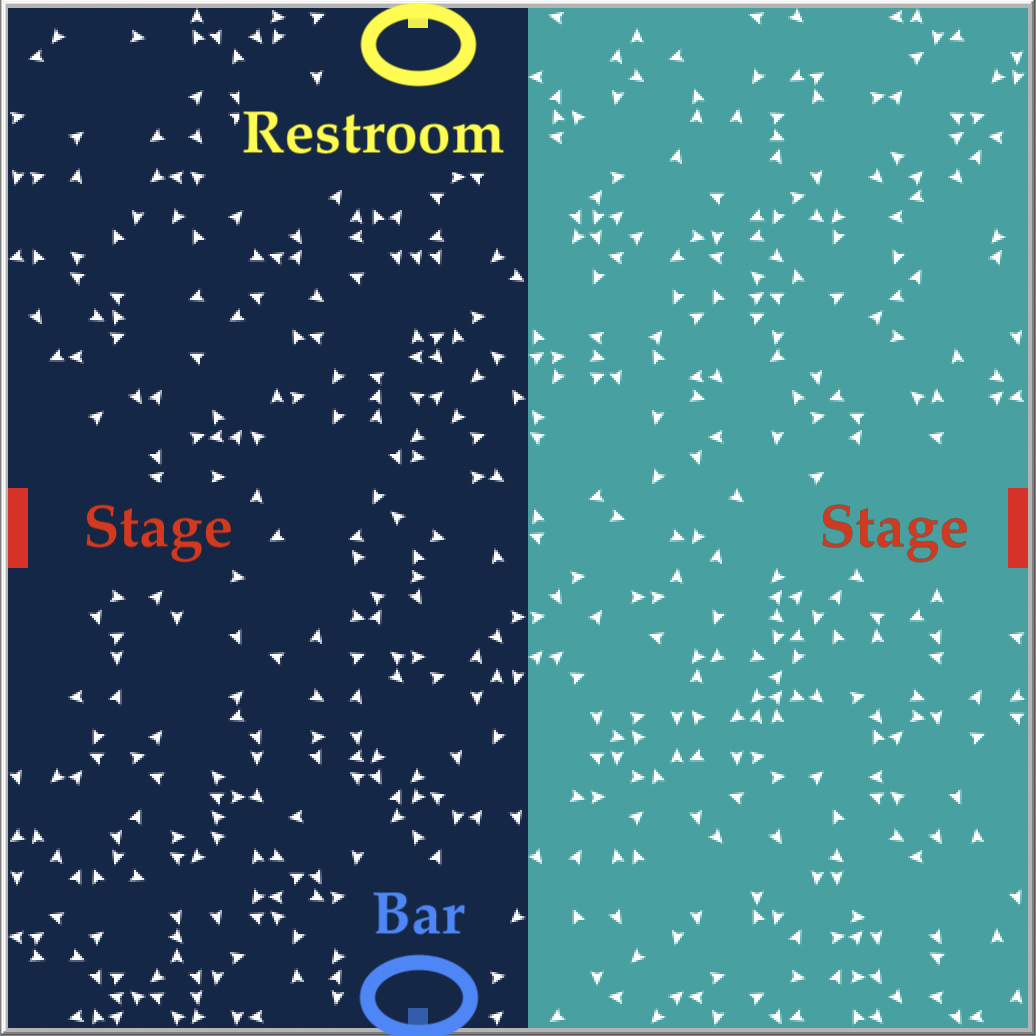}
         \caption{Map A: The frequency of stage switching (F) and Average Panic/Surge (APS) are $0.010$ and $0.85$ respectively.}
         \label{fig:mapa}
     \end{subfigure}
     \hfill
     \begin{subfigure}[b]{0.3\textwidth}
         \centering
         \includegraphics[width=\textwidth]{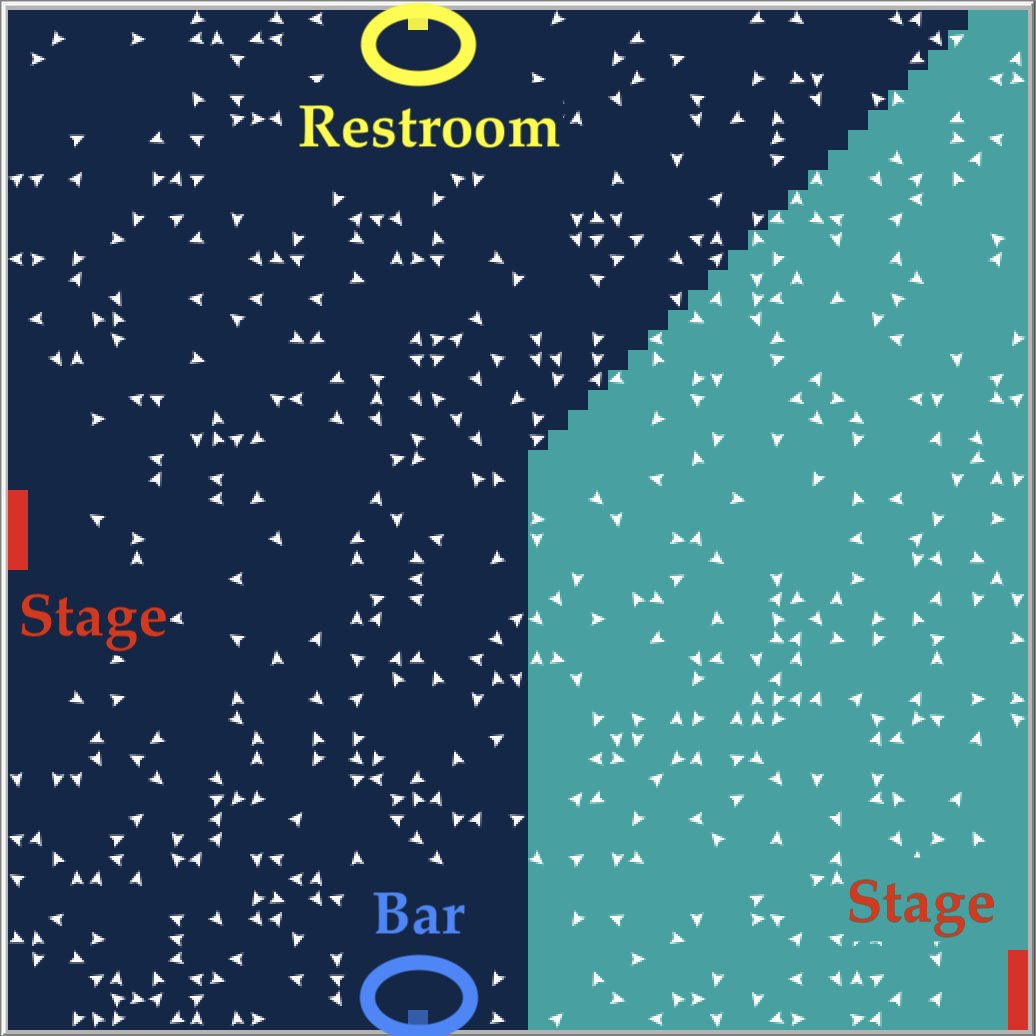}
         \caption{Map B: The frequency of stage switching (F) and Average Panic/Surge (APS) are $0.009$ and $1.05$ respectively.}
         \label{fig:mapb}
     \end{subfigure}
     \hfill
     \begin{subfigure}[b]{0.3\textwidth}
         \centering
         \includegraphics[width=\textwidth]{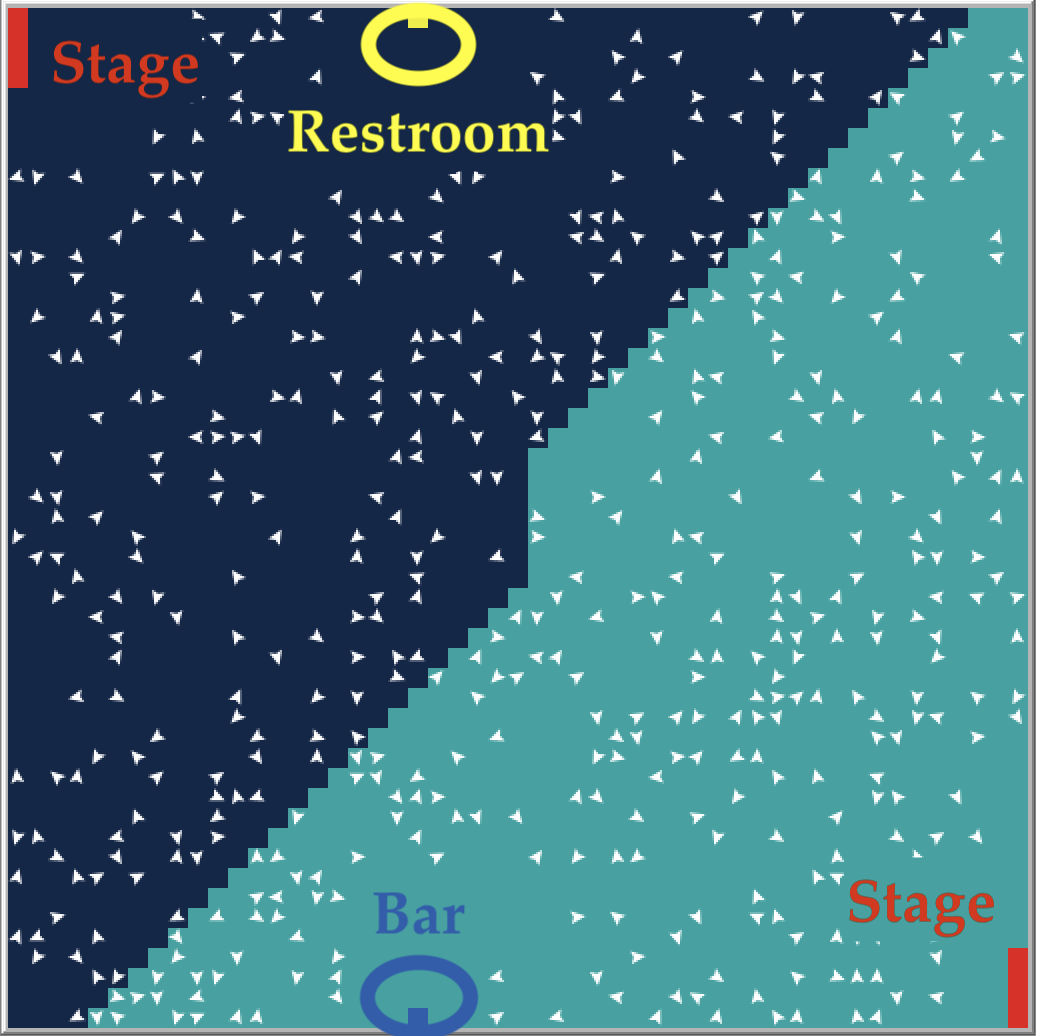}
         \caption{Map C: The frequency of stage switching (F) and Average Panic/Surge (APS) are $0.007$ and $0.62$ respectively.}
         \label{fig:mapc}
     \end{subfigure}
        \caption{\small{Three different stage setups.}}\vspace{-2mm}
        \label{fig:3stages}
\end{figure*}